\documentclass[aps,prb,twocolumn,groupedaddress,amsmath,amssymb]{revtex4}

\usepackage{amssymb}
\usepackage{amsmath}
\usepackage{graphicx}
\usepackage{subfigure}
\usepackage{textcomp}
\usepackage{color}
\usepackage{amsfonts}
\usepackage{bbold}
\usepackage{dsfont}
\usepackage{epsfig}
\usepackage{hyperref}

\newcommand{\arctanh}[1]{\operatorname{arctan}}

\begin{document}

\title{Monte-Carlo phase diagram of a Hubbard-Peierls model in the search for spin crossover transition in $\pi$-conjugated polymers}

\author{S.~Bhattacharya,  M.S.~Ferreira and S.~Sanvito}
\affiliation{School of Physics and CRANN, Trinity College, Dublin 2, Ireland}

\date{\today}

\begin{abstract}
\vspace{0.5cm}
We present a Monte Carlo study of the finite temperature properties of an extended Hubbard-Peierls model describing one 
dimensional $\pi$-conjugated polymers. The model incorporates electron-phonon and hyperfine interaction and it is solved 
at the mean field level for half filling. In particular we explore the model as a function of the strength of electron-electron and 
electron-phonon interactions. At low temperature the system presents a diamagnetic to antiferromagnetic transition as the electron-electron 
interaction strength increases. At the same time by increasing the electron-phonon coupling there is a transition from a homogeneous 
to a Peierls dimerized geometry. As expected such a Peierls dimerized phase disappears at finite temperature as a result of 
thermal vibrations. More intriguing is the interplay between the electron-phonon and the electron-electron interactions at finite 
temperature. In particular we demonstrate that for a certain region of the parameter space there is a spin-crossover, where 
the system transits from a low-spin to a high-spin state as the temperature increases. In close analogy to standard spin-crossover 
in divalent magnetic molecules such a transition is entropy driven. Finally we discuss the r\^ole played by the hyperfine interaction 
over the phase diagram.       
\end{abstract}

\maketitle


\section{Introduction}

Over the past few years the field of organic spintronics has received interest from a growing and diverse scientific community 
\cite{Alex,Stefano,Stefano2,dediu1,WW1}. This is primarily driven by its potential for opening new avenues to cheap, low-weight, 
mechanically flexible, chemically inert and bottom up fabricated spin-devices. The most important advantage of organic materials
\cite{Greg,harris,VIK,sbandy} over their inorganic counterparts \cite{das} is the strong suppression of any efficient spin flipping 
mechanisms, resulting in extremely long spin lifetimes. Spin orbit coupling is very weak in organic semiconductors, 
as Carbon has a low atomic number ($Z=12$) and the strength of spin orbit interaction is proportional to $Z^{4}$. Furthermore
also the hyperfine interaction is weak in molecules, since there are very few nuclei with non-vanishing magnetic moment in the 
upper part of the periodic table and the molecular orbitals typically responsible for the electron transport are extended 
$\pi$-type \cite{Alex2}. A recent experimental work \cite{fert} on 
(La,Sr)MnO$_{3}$(LSMO)/\textit{tris}[8-hydroxyquinoline] Aluminum(Alq$_{3}$)/Co 
tunnel junctions reported a giant tunnel magneto resistance of up to 300\% at low temperatures. This is a value that compares 
well with the best inorganic tunnel junctions\cite{ikeda} and it is attributed to a favorable interaction between the organic molecules
and the magnetic metallic surface \cite{fert,spinterface}.

Amongst the various possible materials for organic spintronics, $\pi$-conjugated organic semiconductors appears very 
appealing \cite{dediu2,vardeny}. This is because of their very long and relatively temperature independent spin relaxation time
and their ability to form good interfaces with metal electrodes when incorporated in spin-valve-like devices. The most relevant 
structural feature of such polymers is their planar shape. The $\pi$-electron wave function distends from the molecular plane 
and it can easily interact with the wave function of adjacent molecules. Therefore a face-to-face molecular configuration is 
usually stabilized through strong $\pi$-$\pi$ bonding and diffusive Van der Waals interaction. This leads to a molecular stacking 
arrangement resulting in the formation of a low dimensional lattice. Because of this peculiar structural conformation many of
the $\pi$-conjugated polymers can be described by simplified one dimensional (1D) model Hamiltonians \cite{picozzi}.

An important aspect for the successful integration of organic semiconductors in magnetic memories and in magnetic switching 
devices is the feasibility of manipulating the spin orientation in the organic media. This is difficult to achieve in non-magnetic 
molecules by the tiny non-equilibrium spin population originating from spin injection. In fact the standard techniques 
for manipulation used in inorganic semiconductors, for example optical methods, are ineffective because of the weak spin-orbit 
interaction. A more promising option is that of manipulating the internal spin degrees of freedom of the organic medium, when 
this is magnetic. Intriguingly there is a vast class of molecules, generally known as spin crossover compounds, whose
spin state can be changed from low spin to high spin by an external perturbation \cite{SC}. Since the crossover transition 
is entropy driven, it is most typically achieved by increasing the temperature, although also light, pressure and electro-chemical 
redox reactions can all produce it. Most recently the possibility of spin crossover driven by static electric fields has been proposed 
theoretically \cite{Nadjib,Kim,Andrea}. Notably all these processes occur in compounds incorporating a transition metal 
[usually Fe(II)], which is responsible for the magnetic moment. Thus, it is interesting to explore the possibility of obtaining
spin crossover in organic $\pi$-conjugated polymers. 

In this work we use a combination of energy minimization techniques and Monte Carlo (MC) simulations to investigate the 
temperature dependent phase diagrams for a model Hamiltonian describing $\pi$-conjugated polymers. In particular, we 
explore the parameter space of the model in the search of a spin crossover transition. Intriguingly we find that the interplay
between on-site Coulomb repulsion and electron-phonon (el-ph) coupling, leading to Peierls distortion, can be responsible for
spin crossover.

\section{Computational Methods}

We consider an extended single-site Hubbard-Peierls model \cite{Hubbard} for a 1D lattice described by the following Hamiltonian
\begin{eqnarray}\label{hamiltonian}
\textit{H}&=&\sum_{i\sigma}{\epsilon_{i\sigma} c_{i\sigma}^\dagger c_{i\sigma}}+ \nonumber \\
&+&\sum\limits_{ij}{[t_{ij} +\alpha (q_{i}-q_{j})] (c_{i\sigma}^\dagger c_{j\sigma} + c_{j\sigma}^\dagger c_{i\sigma})} \nonumber + \nonumber \\ 
&+& U \sum\limits_{i} {\hat{n_{i \uparrow}} \hat{n_{i \downarrow}}} + J_\mathrm{H} \sum\limits_{i\alpha\beta} {\vec{S}_{i}\cdot[c_{i\alpha}^{\dagger}  (\vec{\sigma}_{\alpha \beta}) c_{i\beta} ] }+\nonumber \\ 
&+& \frac{1}{2} k \sum\limits_{i,j} {(q_{i}-q_{j})^2}
\end{eqnarray}
\noindent where $c_{i\sigma}^\dagger$($c_{i\sigma}$) denotes the creation (annihilation) operator for an electron at site $i$ with 
spin $\sigma$, $\epsilon_{i\sigma}=\epsilon$ is the onsite energy and $t_{ij}$ is the transfer integral for an uniform undimerized 
lattice. Here we consider only nearest neighbour hopping, i.e. $t_{ij}=t$ for $i=j\pm 1$ and $t_{ij}=0$ for any other $(i,j)$ pair. The other microscopic 
parameters of the model are the el-ph coupling parameter, $\alpha$, the Hubbard repulsion strength, $U$, the hyperfine 
exchange $J_\mathrm{H}$, and the elastic constant, $k$. Thus, the second term of the Hamiltonian, in addition to electron 
hopping, describes the el-ph coupling, with $q_{i}$ being the atomic displacement of site $i$ (we consider 1D longitudinal motion only). 
The third term is the standard on-site Hubbard repulsion while the fourth one describes a Heisenberg-like interaction between the electron 
spins, $c_{i\alpha}^{\dagger}  (\vec{\sigma}_{\alpha \beta}) c_{i\beta}$, and a set of classical vectors, $\vec{S}_i$, representing the nuclear
spins. Finally the last term is classical elastic energy. 

\begin{figure*}[ht]
\centerline{\epsfxsize=7.5cm\epsffile{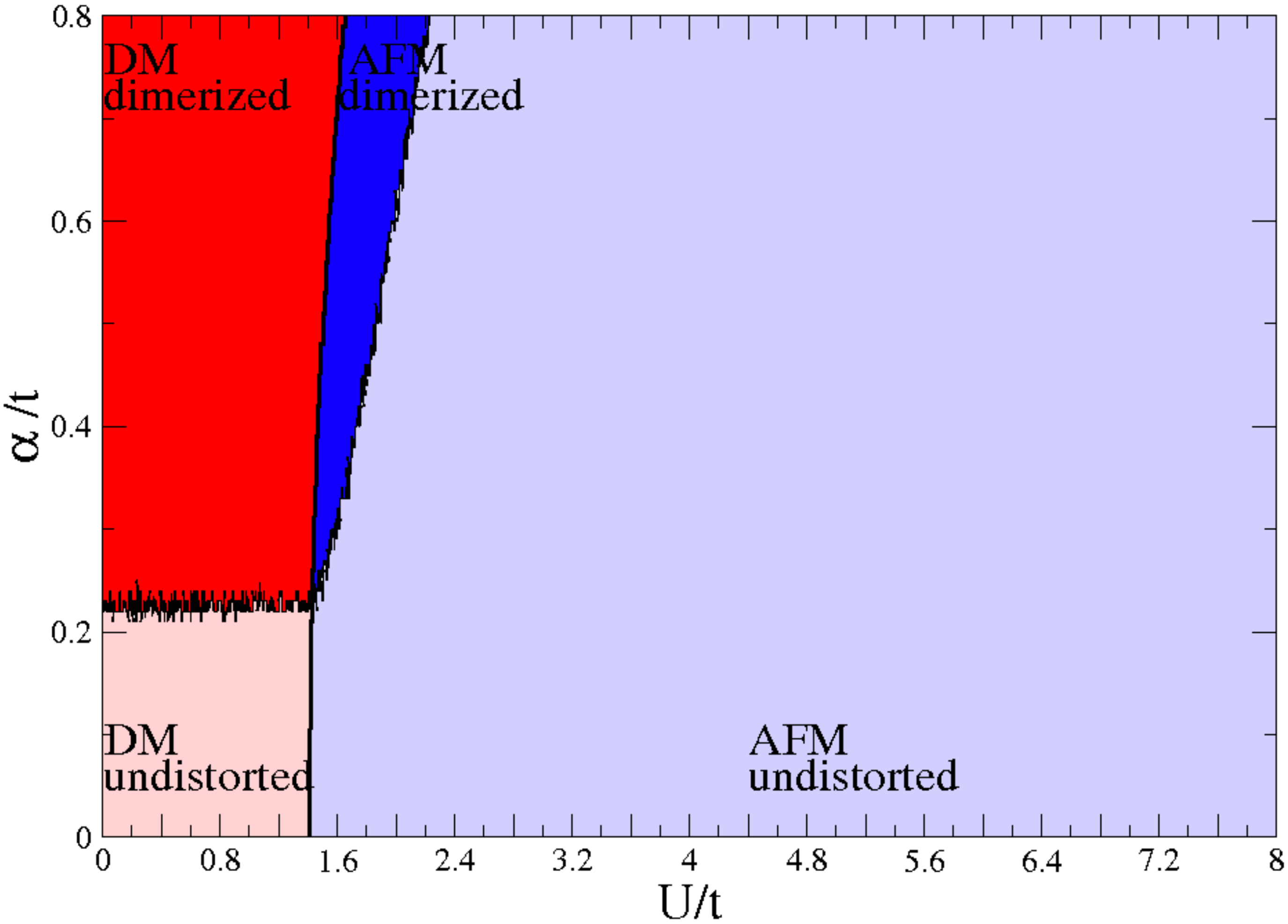}\hspace{0.5cm}\epsfxsize=8.0cm\epsffile{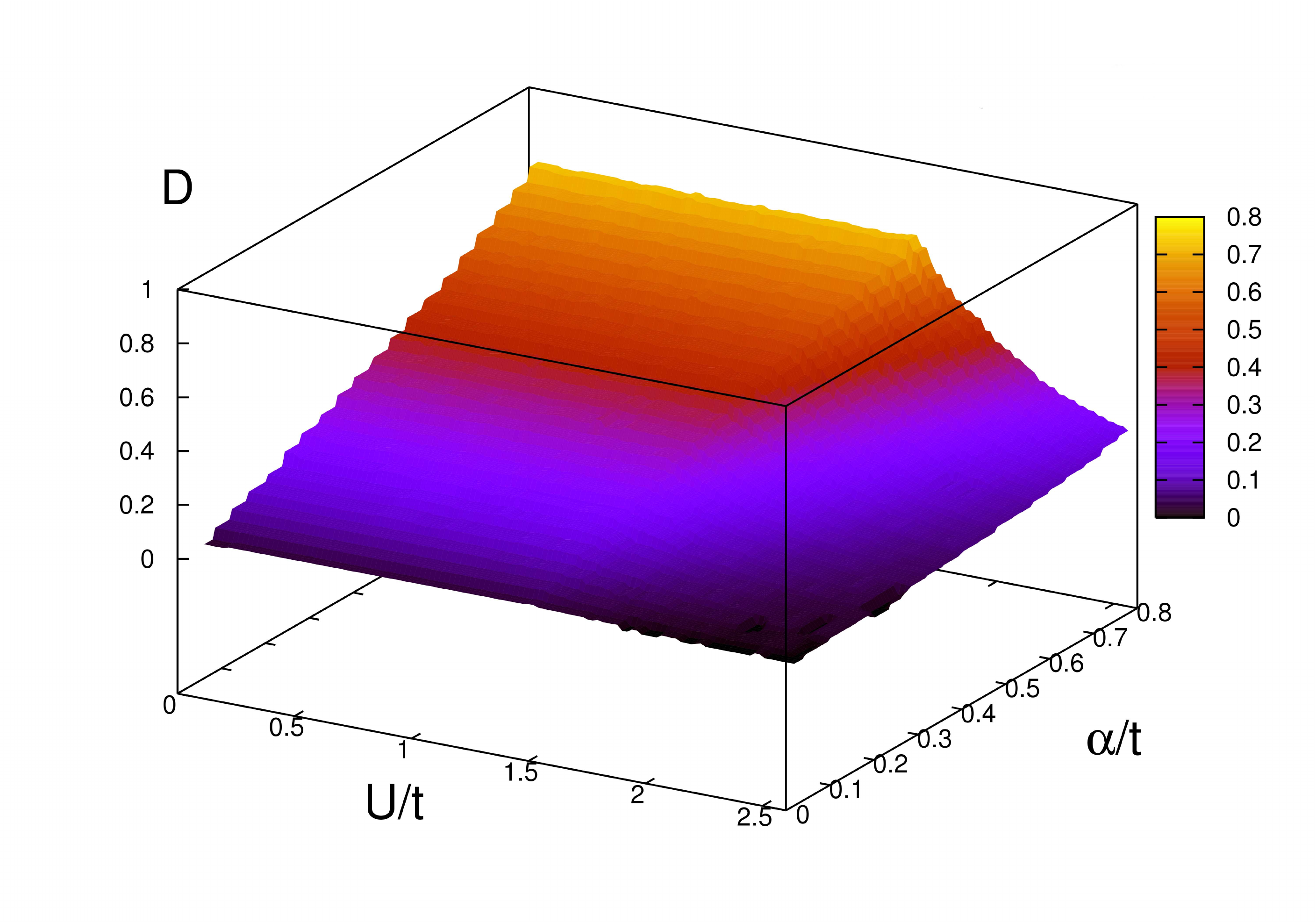}}
\caption{(Color online) $\alpha$-$U$ phase diagram at $T=0$ for the model 1D polymers investigated. In the left panel we present
the schematic phase diagram, while in the right panel the dimerization parameter, $D$, as a function of $\alpha/t$ and $U/t$.}
\label{groundstate}
\end{figure*}

In all our calculations we consider only the half-filling case (one electron per site) so that the model exhibits an insulating behaviour at 
$T=0$ for an undistorted infinite chain. The data presented here are for 10 site long chains as further tests for longer chains (20 sites)
gave rather similar results. Our approach consists in replacing the Hubbard term with its unrestricted mean-field 
approximation \cite{meanfield} and then in solving the Hamiltonian self-consistently for different lattice displacements $\{q_{i}\}$. 
Energy minimization is performed by conjugate gradient over $\{q_{i}\}$ and further verified by additional simulated annealing \cite{SA}. 
The bond lengths, $x_{i}$, are calculated from the ground state displacements as $x_{i}=d+q_{i+1}-q_{i}$, where $d$ is the equillibrium 
bond distance. The main observables calculated are the dimerization parameter, $D$, and the local magnetic moment per site, $m_{i}$. 
These are defined respectively in Eq.~\eqref{dim} and Eq.~\eqref{moment} 
\begin{equation}\label{dim}
D = \displaystyle\sum\limits_{i=1}^{N-2}  |x_{i}-x_{i+1}|+|x_{i+1}-x_{i+2}|\:,
\end{equation}
\begin{equation}\label{moment}
m_{i} = |n_{i}^{\uparrow}-n_{i}^{\downarrow}|\:.
\end{equation}
Note that in Eq.~\eqref{moment} $n_{i}^{\sigma}$ is the electron 
spin density at site $i$, which is calculated as the diagonal matrix elements of the density matrix 
$\hat{\rho}=\sum\limits_{n=1}^\mathrm{occupied}\{|\Psi_{n} \rangle \langle \Psi_{n}|\tfrac{1}{\mathrm{e}^{\beta E_{n}}-1} \}$, 
where $E_{n}$ and $|\Psi_{n}\rangle$ are respectively $n$-th electronic eigenvalues and eigenvectors.
Our results are then plotted in a phase-diagram like form, where the different phases are presented as a function of $\alpha$ and $U$. 

For finite temperature simulations we consider the system described by the Hamiltonian of Eq.~\eqref{hamiltonian} and 
evolve the classical dynamical variables $\{q_{i}\}$ and $\{\vec{S}_i\}$ by using the standard Metropolis algorithm 
\cite{metropolis}. Note that since $J_\mathrm{H}\ll t$ the hyperfine interaction has little effect on the $\alpha$-$U$ phase
diagram, so that in what follows we will neglect it unless otherwise indicated. As such the only classical dynamical variables 
are the atomic displacements $\{q_{i}\}$. In the Metropolis algorithm the acceptance probablity of a new state is unity if the new 
configuration has an energy lower than that of the old configuration. Otherwise it is given by the Boltzmann factor 
$\mathrm{e}^{-\frac{ \Delta G}{k_\mathrm{B} T }}$, where $ \Delta G $ is the difference in the Gibbs free energy between the old and 
new configuration. By using the grand-canonical ensemble the Gibbs free energy $G$ can be written as
\begin{equation}\label{gee}
G (\{q_{i}\})=-\dfrac{1}{\beta}\sum_{n=1}^{N_{o}} \ln\left(1+\mathrm{e}^{\:-\beta[E_{n}(\{q_{i}\})-\mu]}\right)\:,
\end{equation}
where the chemical potential $\mu$ is obtained from
\begin{equation}
N = \sum_{n} \dfrac{1}{\mathrm{e}^{\beta[E_{n}(\{q_{i}\})-\mu]}+1}\:,
\end{equation}
with $\beta=\tfrac{1}{k_\mathrm{B}T}$ being the inverse temperature and $N$ the total number of electrons ($N=10$ here). 
For every value of ($\alpha/t$, $U/t$) and each temperature the system is allowed to reach equilibrium. Then both $D$ 
and $m_i$ are calculated over several million MC steps. 

In what follows we will express all energy related quantities (including the temperature) as a function of $t$, which sets the energy scale
of the problem. The on-site energy is taken to be zero and $k$ is 5~$t$/\AA$^2$. Note that for $t\sim$2.5~eV this corresponds to 
$k=12.5$~eV/\AA$^2$, which is in between the value for the H$_2$ molecule and that of Au monoatomic chains \cite{Will}. Also it
is important to note that $k>5t/$\AA$^2$ is a value commonly used in recent literature about transport in organic polymers \cite{picozzi,saxena}.

\section{Results and Discussion}

\subsection{Ground State}

Let us begin our discussion by investigating the $\alpha$-$U$ phase diagram at $T=0$, which is presented in Fig.~\ref{groundstate}. 
This is populated by four difference phases characterized by the different combined structural and magnetic properties of the chain. 
In particular there are two magnetic states and two geometrical configurations. For small $\alpha/t$ and $U/t$ the chains are 
undistorted (the atomic spacing is approximately uniform throughout the chain) and in a non-spin-polarized diamagnetic (DM) state. 
We denote this phase as DM-undistorted. As $U/t$ increases eventually a spin-polarized solution develops. This is the one expected 
from the mean field Hubbard model at half-filling, i.e. it is a antiferromagnetic (AFM) phase, where local moments form at each atomic 
site, but their orientation alternates along the chain. Such an AFM phase may or may not be accompanied by a structural distortion, 
depending on the value of $\alpha/t$. In general however there is a vast region of the model parameter space, where no significant 
distortion appears for the AFM spin state. We denote this phase as AFM-undistorted. 

As $\alpha/t$ increases for moderate $U/t$ the system progressively developed a Peierls instability and makes a transition to a geometry
where long bond distances alternate to short ones. This dimerized phase, expected for the non-interacting case, remains diamagnetic 
(DM-dimerized) for small $U/t$ but can coexist with a AFM solution for a significant range of parameter. In summary the phase diagram is
characterized by a competition between the on-site repulsion, driving the magnetic instability, and the el-ph coupling, driving the 
Peierls distortion. 

In the discussion we have assigned the phase boundary of the magnetic transition to the condition $\sum |m_i|\ne0$. In contrast assigning 
the phase boundary to the Peierls transition is more complicated since $D$ changes continuously upon increasing $\alpha$. Thus we have 
used the operational definition of placing the phase boundaries at $D=0.17$, which is interpreted as representative of strong dimerization. 
The complete evolution of $D$ as a function of $U/t$ and $\alpha/t$ is presented in the right hand side panel of Fig.~\ref{groundstate}.
The figure clearly reveals the interplay between el-ph coupling and Coulomb repulsion. In fact $D$ grows almost linearly with $\alpha/t$
for small $U/t$ but then is drastically reduced as $U/t$ grows.

\subsection{Finite temperature phase diagram}

\begin{figure*}[htbp]
\centerline{\epsfxsize=5.8cm\epsffile{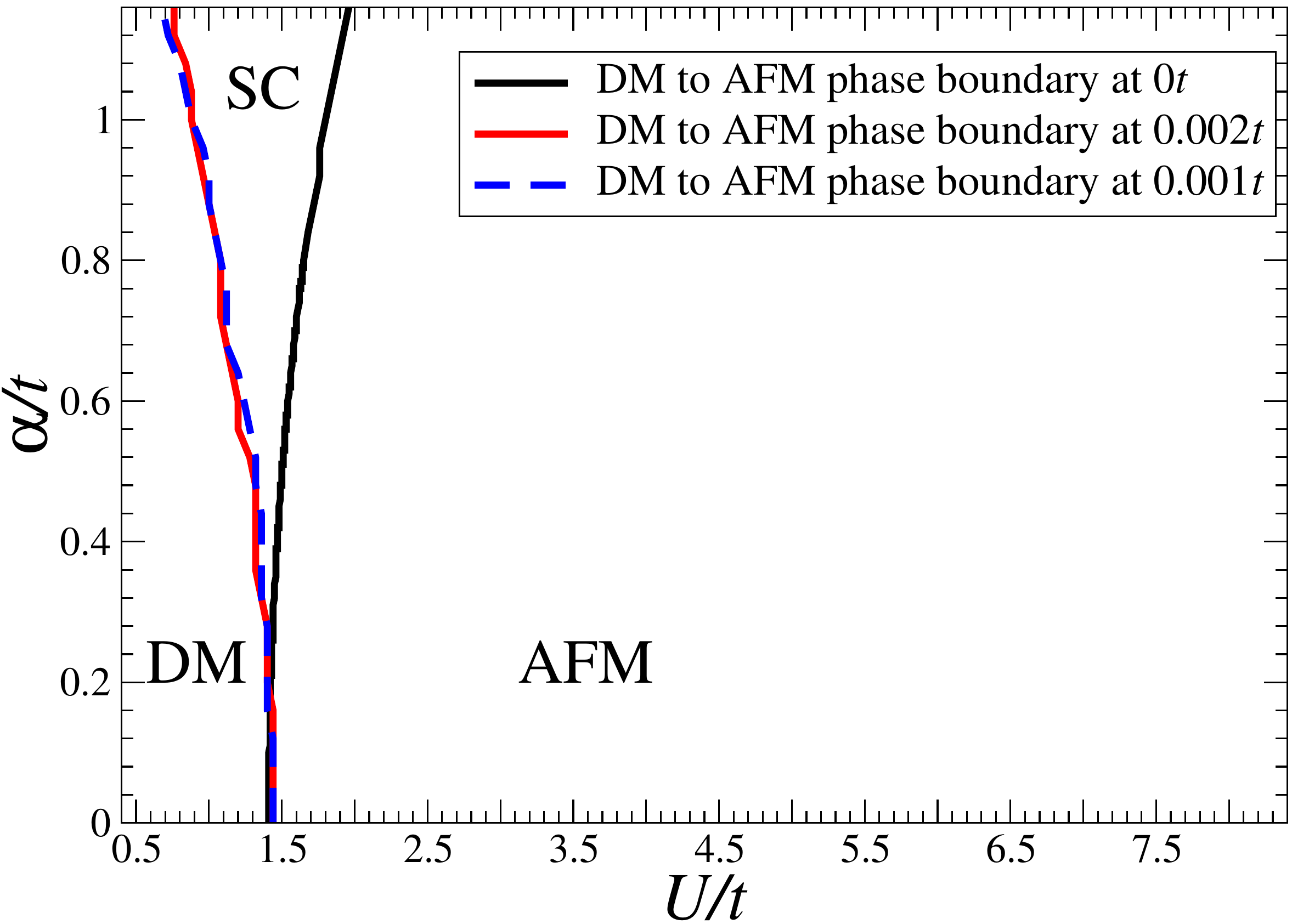}\epsfxsize=6.3cm\epsffile{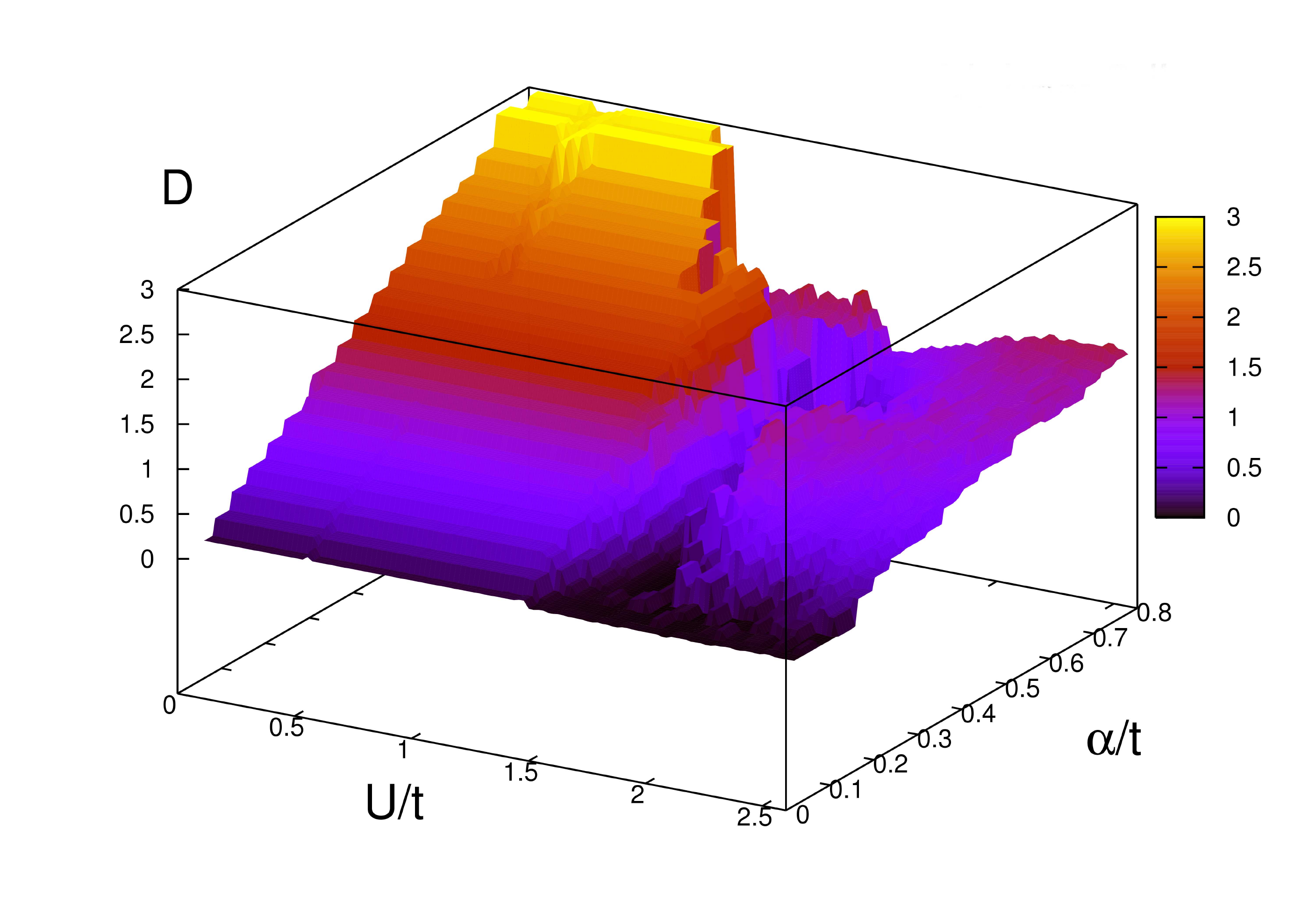}\epsfxsize=6.3cm\epsffile{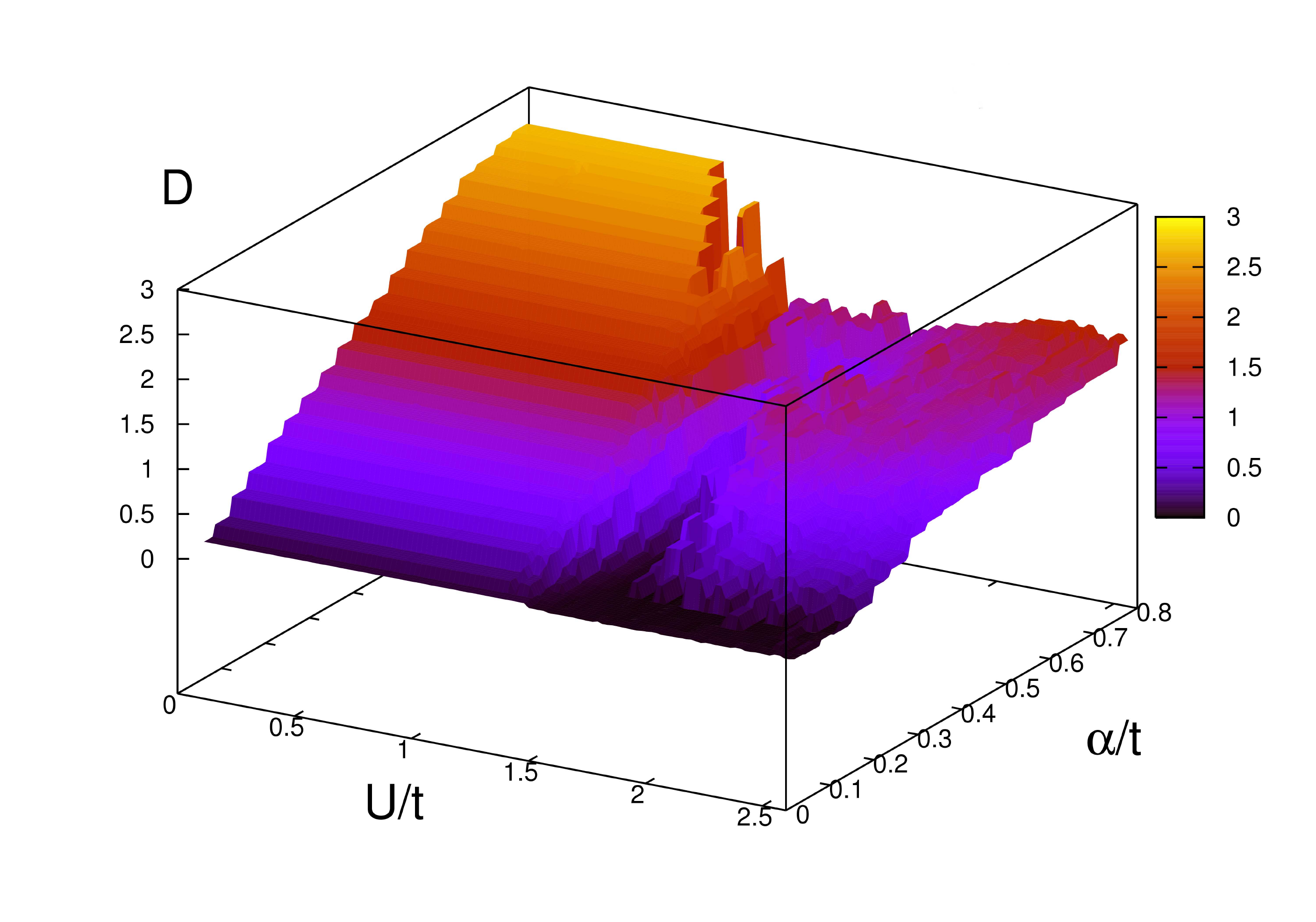}}
\caption{(Color online) Finite temperature $\alpha$-$U$ phase diagram for the model 1D polymers investigated. In the left panel we present
the schematic phase diagram for two different temperatures, respectively $T=0.001\:t$ and $T=0.002\:t$. Note that now there are only two 
phases (DM and AFM) and there is no longer a distorted (Peierls) geometrical configuration. In the picture we also report the DM-AFM 
phase boundary for the $T=0$ case. Thus the region comprised between the solid black line either the solid red of the dashed blue line
is characterized by a DM to AFM phase transition (spin crossover - SC - region) as a function of temperature. In the middle ($T=0.001\:t$) and the 
right-hand side panel ($T=0.002\:t$) we present the parameter $D$ as a function of $\alpha/t$ and $U/t$ which indicates the degree of disorder 
in the bond distances at finite temperature. }
\label{MC}
\end{figure*}

We now move to study the finite temperature properties of our model. These are summarized in Fig.~\ref{MC}, where we present the 
$\alpha$-$U$ phase diagram for two representative temperatures, respectively $\beta=200, 400$~1/eV (these correspond respectively 
to $T=0.002\:t, 0.001\:t$ or, for $t=2.5$~eV, to $T=58, 29$~K).
The most important feature of the finite temperature plots is the complete absence of a structural phase transition. This means that in general 
the system does not dimerize any longer as the temperature is increased. The dimerization is instead replaced by a general increase in bond
length and by a random distribution of the various bonds along the chain.

More details about the structure of the chains at finite temperature can be found in the right-hand side panes of Fig.~\ref{MC}, which show the 
quantity $D$ [defined in Eq.~\eqref{dim}] as a function of $U/t$ and $\alpha/t$ for $T=0.001t$ and $0.002t$. Note that at finite temperature the 
quantity $D$ plotted is a measure of the disorder in the bond distances of the chain as thermal vibrations onset. As such $D$ is maximum 
at high values of the el-ph coupling and low $U/t$ but falls rapidly as $U/t$ increases.

\begin{figure}[ht]
\includegraphics[width=7cm, clip=true]{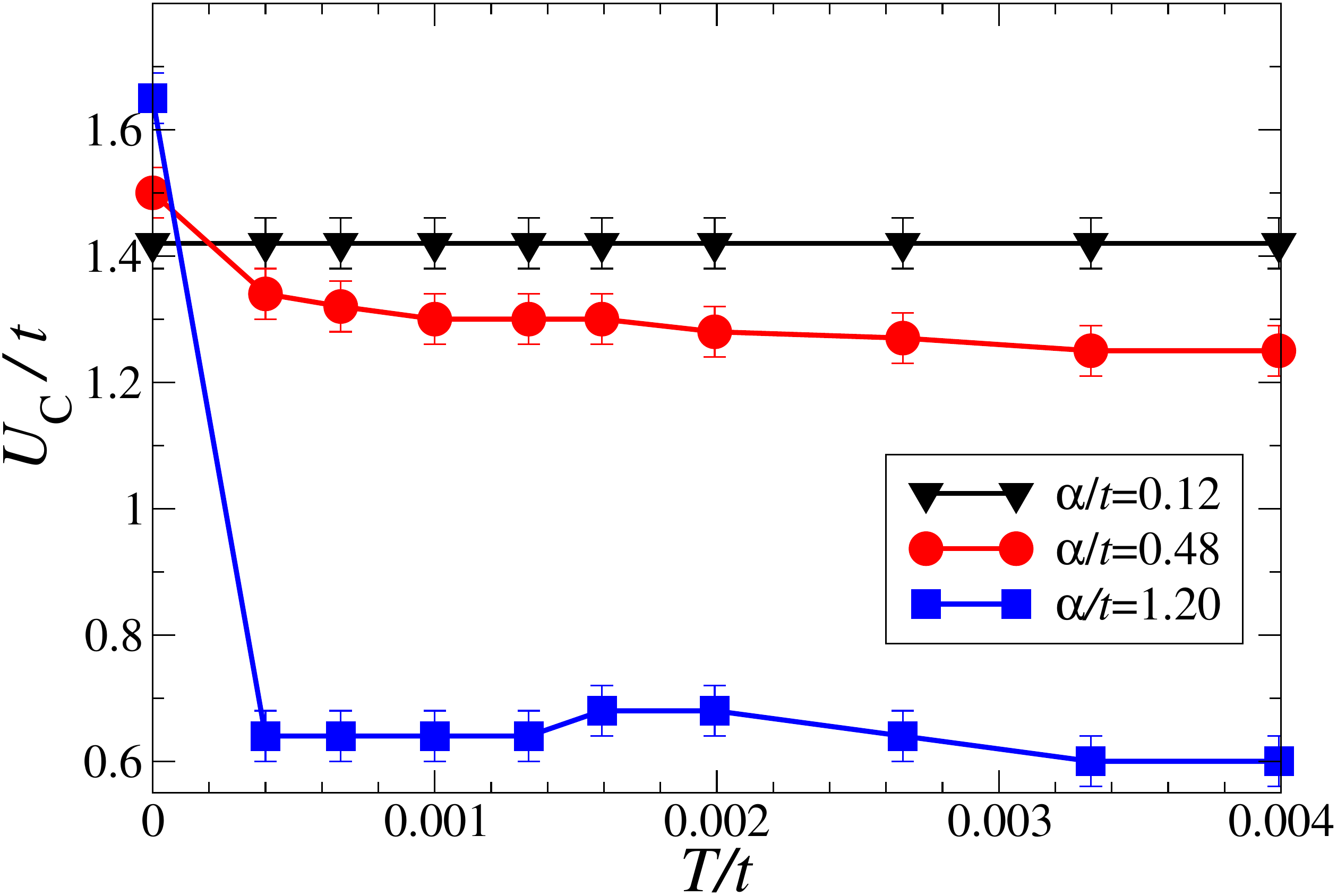}
\caption{(Color online) Critical on-site Coulomb repulsion, $U_\mathrm{C}$, needed for a magnetic solution as a function of temperature 
and for three different values of the el-ph coupling strength. The error bars correspond to spacing of the interpolation grid used to 
extract $U_\mathrm{C}$.}\label{boundary}
\end{figure}

The second most striking feature of the finite temperature $\alpha$-$U$ phase diagram is the movement of the DM-AFM phase boundary 
towards lower $U/t$ as the temperature increases. This essentially means that as the temperature grows it takes less on-site Coulomb
interaction to drive the system towards a magnetic instability. We further explore this finding in figure~\ref{boundary}, where we present
the critical $U$ value, $U_\mathrm{C}$, at which the magnetic phase develops. This effectively represents the position of the DM-AFM phase 
boundary. In particular $U_\mathrm{C}$ is plotted as a function of the temperature, $T$, and for three different el-ph strengths $\alpha/t$.

In general we find that the DM-AFM phase boundary moves in response to the disappearing of the distorted phase. Thus for the 
lower value of $\alpha/t$ (0.12), for which there is no distorted phase even at $T=0$, the phase boundary does not change as the temperature
is increased and $U_\mathrm{C}$ remains constant at $\sim1.4\:t$. For the larger values of the el-ph interaction strengths investigated the
$T=0$ phase diagram presents both a distorted and a homogeneous structural phase depending on $U$. In this case the DM-AFM
phase boundary (i.e. $U_\mathrm{C}$) decreases fast at low temperatures in response to the melting of the distorted phase and then becomes 
essentially constant. 

An important consequence of the movement of the DM-AFM phase boundary as the temperature increases is the fact that there is 
a vast region in the $\alpha$-$U$ diagram where the system makes a DM to AFM transition as the temperature increases. Such a
region is the one enclosed between the two vertical lines marking respectively the phase boundary at finite temperature and at
$T=0$ in figure~\ref{MC}. For the particular values of $\alpha$ and $U$ found in such a region (called the spin crossover -SC- region)
there is a temperature driven spin crossover. This is analyzed next. 

\subsection{Spin crossover}

We now explain the spin crossover transition by using the standard framework of spin crossover usually employed for magnetic
molecules incorporating divalent transition metals \cite{SC}. In general the thermodynamically stable phase at finite temperature of a 
system that can assume different competing configurations is the one with the lowest Gibbs free energy, $G$. For the present
case, where the competition is among the DM and the AFM phases, the relevant quantity is the difference, 
$\Delta G=G_\mathrm{AFM}-G_\mathrm{DM}$, between their Gibbs energies, 
\begin{equation}\label{gibbs}
\Delta G=\Delta H-T\Delta S  
\end{equation}
where $\Delta H=H_\mathrm{AFM}-H_\mathrm{DM}$ and $\Delta S=S_\mathrm{AFM}-S_\mathrm{DM}$ are respective the enthalpy and 
entropy differences. 

For standard spin crossover $\Delta H >0$, so that the most stable phase at low temperature in low spin (DM here). However, since the 
crossover transition is associated to the softening of the phonon modes of the first coordination shell around the transition metal 
and to the formation of a local magnetic moment, we also have $\Delta S>0$. Hence as the temperature increases the entropic 
contribution to the Gibbs energy may eventually dominate over the enthalpic one and drive a phase transition. We now want to
establish that the same mechanism holds for the spin crossover region of the $\alpha$-$U$ diagram of Fig.~\ref{MC}.

\begin{figure}[ht]
\includegraphics[width=8cm, clip=true]{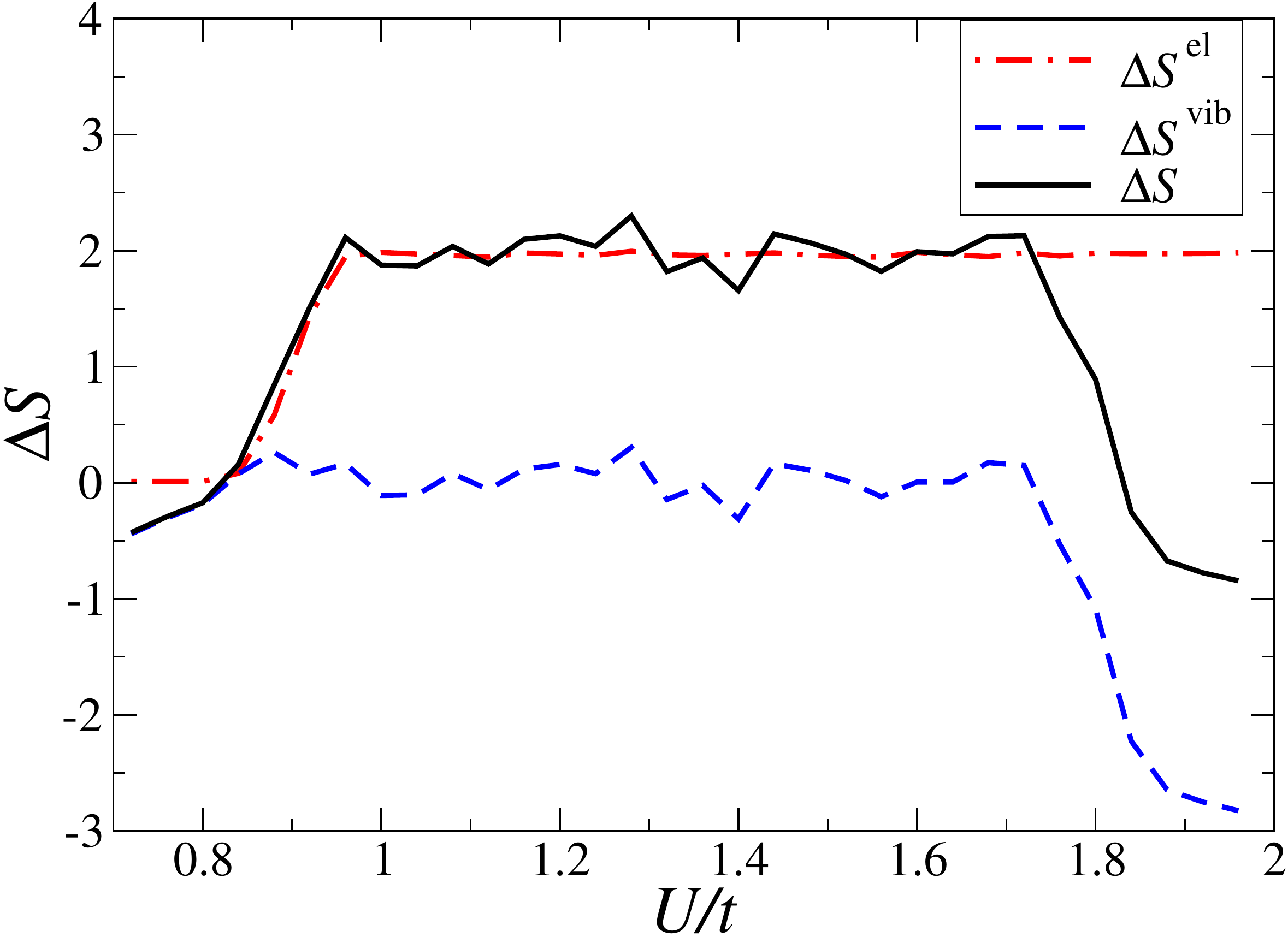}
\caption{(Color online) Entropy difference, $\Delta S$, between the DM and AFM phases. The individual electronic, $\Delta S^\mathrm{\:el}$,
and vibrational, $\Delta S^\mathrm{\:vib}$, contributions as well as their sum, $\Delta S$, are presented as a function of $U/t$ for $\alpha=1\:t$. 
The entropy is presented in adimensional units where both the energy and the temperature are in units of the hopping integral $t$.
Note that there is a large region $0.84t<U<1.85t$, where $\Delta S>0$. In this region the entropic contribution to the Gibbs free energy can drive
a spin crossover transition.}\label{entropy}
\end{figure}

We have already demonstrated (see figure~\ref{groundstate}) that for $T=0$ the spin crossover region is occupied by the DM phase, 
meaning that $\Delta H >0$. Therefore one has only left to show that also $\Delta S>0$. In general the entropy comprises of two main 
contributions, an electronic, $S^\mathrm{\:el}$, and a vibrational one, $S^\mathrm{\:vib}$. Since the AFM phase is characterized by 
local spins, which are absent for the DM phase, we can immediately conclude that 
$\Delta S^\mathrm{\:el}=S^\mathrm{\:el}_\mathrm{AFM}-S^\mathrm{\:el}_\mathrm{DM}>0$. A more precise evaluation of $S^\mathrm{\:el}$
can be obtained by computing 
\begin{equation}\label{S_el}
S^\mathrm{\:el}= -k_\mathrm{B} \mathrm{Tr}[\hat{\rho} \ln \hat{\rho}]\:,
\end{equation}
where $\hat{\rho}$ is the system density matrix. The calculated $\Delta S^\mathrm{\:el}$ as a function of $U/t$ for the representative 
value of $\alpha/t=1$, for which the spin crossover region is quite large, is presented in figure~\ref{entropy}. The electron densities 
entering the evaluation of $S^\mathrm{\:el}$ have been calculated as follows. For the low temperature phase (DM) $\rho$ is calculated
by fixing the occupation to $n_i=1/2$ for every site and the geometry is that obtained from the $T=0$ diagram. In contrast the entropy of the 
high-temperature phase is computed from a density matrix in which the occupation is fixed to the proper antiferromagnetic state (the temperature
is $T=0.001t$) and the geometry is again that of the $T=0$ solution. We have checked that the finite temperature geometry is rather similar
to that obtained at $T=0$ for such a density-constraint solution. 

The general trend can be understood as follows. For small values of $U/t$ there is no local magnetic moment formation, regardless 
of the chain geometry, so that $\Delta S^\mathrm{\:el} = 0$. Then, as $U/t$ gets larger a local magnetic moment gradually appears at each site, producing
a linear increase of $\Delta S^\mathrm{\:el}$. Such an increases continues until the local moment reaches the maximum value compatible with the chosen
electron filling, which is 1~$\mu_\mathrm{B}$ ($\mu_\mathrm{B}$ is the Bohr magneton) for half-filling. At this point there is no further change
in the electronic entropy and $\Delta S^\mathrm{\:el}$ saturates to a positive value. 

Similarly we can also estimate the vibrational contribution to the entropy. This is obtained from the molecule phonon spectrum, 
$\hbar\omega_i$, as
\begin{equation}\label{S_vib}
S^\mathrm{\:vib}= k_\mathrm{B} \ln \sum\limits_{i}^{N_{v}} \frac{1}{e^{\beta \hbar \omega_{i}}-1} + k_\mathrm{B}\beta \sum\limits_{i}^{N_{v}} \hbar \omega_{i}\:,
\end{equation}
where $\omega_i$ is the vibration frequency of the $i$-th mode and $N_v$ is the total number of modes. We calculate the phonon spectra 
of the AFM and DM configurations by diagonalizing the associated dynamical matrices. These are constructed by finite difference, i.e. by 
displacing the atomic sites by a small fraction of the equilibrium bond length (0.1~\%) and then by numerically evaluating the energy
gradient (the force) associated to such a displacement. The density matrices and the initial geometries used to construct the finite difference
dynamical matrices are the same used for calculating the electronic contribution to the entropy. Also in this case $\Delta S^\mathrm{\:vib}$ 
as a function of $U/t$ for $\alpha/t=1$ is presented in figure~\ref{entropy}. 

In general the vibrational contribution to the entropy difference shows only a small dependence on the Coulomb on-site repulsion and
approximately $\Delta S^\mathrm{\:vib}\sim0$. However we also report a relatively sharp decreases as $U/t$ approaches 1.7. This is
value close to $U_\mathrm{C}$ for $T=0$, i.e. corresponds to a region in the parameter space where our Monte Carlo analysis does not
find a DM-AFM transition and therefore system remains in the AFM state at any of the temperatures investigated. 

In summary the picture emerging from figure~\ref{entropy} is that of a region $0.84t<U<1.85t$ in which $\Delta S$ is always positive. 
This is the only region of the parameter space where the entropy can drive the spin crossover and substantially matches the spin 
crossover region observed in our Monte Carlo simulations (see Fig.~\ref{MC}). We then conclude that also in this case, where the
magnetic moment is not associated to the $d$ shell of a transition metal, the spin crossover is entropy driven. 

Finally before summarizing, we wish to comment on the r\^ole played by the hyperfine interaction. In general we expect very little changes to the 
phase diagram obtained by neglecting the hyperfine contribution to the total energy, since this is rather small for realistic hyperfine coupling 
strengths. In particular we have verified that minor modifications to the $T=0$ phase diagram start to appear for $J_\mathrm{H}$ is the region 
of $0.01\:t$, which corresponds to local magnetic fields of 10$^7$~T (considering $|\vec{S}_i|=1$ and $t$ of the order of 1~eV).

\section{Conclusion}

In summary we have discussed the phase diagram of an extended Hubbard-Peierls model chosen to emulate the electronic structure of 
$\pi-$conjugated polymer chains. The model has been explored both in its $T=0$ ground state and at finite temperature by Monte Carlo
methods combined with a mean field treatment of the Hubbard many-body interaction. At $T=0$ the model presents four different phases 
depending on the relative strength of the Coulomb on-site repulsion $U/t$ and the el-ph coupling strength $\alpha/t$. The four phases
are characterized either by a diamagnetic or an antiferromagnetic magnetic state and by the possible presence of a dimerized geometrical
configuration. By increasing the temperature the structural distortion disappears and only the phase boundary between the DM and the AFM 
solution remains. Intriguingly the position in the $\alpha$-$U$ parameter space of such a phase boundary changes with temperature 
so that there is a region of parameters where a temperature driven DM-AFM spin crossover transition can be found. We have investigated the
nature of the phase transition by calculating the relative entropy of the different magnetic phases and found that this is indeed entropy driven 
as in the most conventional case of divalent magnetic molecules. This suggests that $\pi$-type magnetism can be achieved in organic
polymers and that this can be tuned by temperature. 

\section*{Acknowledgments}

This work is sponsored by Science Foundation of Ireland CSET grant underpinning CRANN. Computational resources have been provided 
by the HEA IITAC project managed by the Trinity Centre for High Performance Computing.


\begin{thebibliography}{99}

\bibitem{Alex}A.R.~Rocha, V. M.~Garc\`ia-Suarez, S.W.~Bailey, CJ.~Lambert, J.~Ferrer and S.~Sanvito, Nature Materials {\bf 4}, 335 (2005).

\bibitem{Stefano}S.~Sanvito, J.~Mat.~Chem. {\bf 17}, 4455 (2007).

\bibitem{Stefano2}S.~Sanvito, Nature Materials {\bf 6}, 803 (2007).

\bibitem{dediu1}V.A.~Dediu , L.E.~Hueso, I.~ Bergenti and C.~Taliani, Nature Materials {\bf 8}, 707 (2009).

\bibitem{WW1}L.~Bogani and W.~Wernsdorfer, Nature Materials {\bf 7}, 179 (2008).

\bibitem{Greg}G.~Szulczewski, S.~Sanvito and J.M.D.~Coey, Nature Materials {\bf 8}, 693 (2009).

\bibitem{harris}C.B.~ Harris,  R.L.~Schlupp and H.~Schuch, Phys. Rev. Lett. {\bf 30}, 1019 (1973).

\bibitem{VIK}V.I.~Krinichnyi , S.D.~Chemerisov and Y.S.~Lebedev, Phys. Rev. B {\bf 55}, 16233 (1997).

\bibitem{sbandy}S.~Pramanik, C.G.~Stefanita, S.~Patibandla, S.~Bandyopadhyay K.~Garre, N.~Harth and  M.~Cahay, 
Nature Nanotech. {\bf 2}, 216 (2007).

\bibitem{das}I.~ Zutic, J.~Fabian and S.~Das Sarma, Rev. Mod. Phys. {\bf 76}, 323 (2004).

\bibitem{Alex2}S.~Sanvito and A.R.~Rocha, J. Comp. Theo. Nano. {\bf 3}, 624 (2006).

\bibitem{fert}C.~Barraud, P.~Seneor, R.~Mattana, S.~Fusil, K.~Bouzehouane, C.~Deranlot, P.~Graziosi, L.~Hueso, I.~Bergenti, V.A.~Dedie, F.~Pertroff and A.~Fert, Nature Phys. {\bf },  (2010).

\bibitem{ikeda}S.~Ikeda, J.~Hayakawa, Y.~Ashizawa, Y. M.~Lee, K.~Miura, H.~Hasegawa, M.~Tsunoda, F.~Matsukura, and 
H.~Ohno, Appl.Phys.Lett. {\bf 95}, 082508 (2008).

\bibitem{spinterface}S.~Sanvito, Nature Physics {\bf 6}, 562 (2010).

\bibitem{dediu2}V.~Dediu, M.~Murgia, F.C.~Matacotta, C.~Taliani and S.~Barbanera, Sol. State Commun. {\bf 122}, 181 (2002).

\bibitem{vardeny}Z. H.~Xiong, D.~Wu, Z. V.~Vardeny, and J.~Shi, Nature (London) {\bf 427}, 821 (2004).

\bibitem{picozzi}G.~Giovannetti, S.~Kumar, A.~Stroppa, J.V.D.~Brink and S.~Picozzi, Phys. Rev. Lett. {\bf 103}, 266401 (2009).

\bibitem{SC}P.~G\"utlich, H.A.~Goodwin, in {\it Spin Crossover in Transition Metal Compounds} (eds. P. G\"utlich, H.A.~Goodwin) (Springer, 2004).

\bibitem{Nadjib}N.~Baadji, M.~Piacenza, T.~Tugsuz, F.~Della Sala, G.~Maruccio, S.~Sanvito, Nature Materials, {\bf 8}, 813 (2009).

\bibitem{Kim} M.~Diefenbach, K.S.~Kim, Angew. Chem. {\bf 119}, 7784 (2007).

\bibitem{Andrea}A.~Droghetti and S.~Sanvito, in preparation.

\bibitem{Hubbard}J.~Hubbard, Proc. Roy. Soc. A {\bf 276}, 238 (1963).

\bibitem{meanfield}P.W.~Anderson in \textit{Fizika Dielectrikon} edited by G.I.~Skanov and K.V.~Filippov 
(Academy of Sciences, Moscow, 1960) p.~290.

\bibitem{SA}S.~Kirkpatrick, C.D.~Gelatt and M.P.~Vecchi, Science {\bf 220}, 4598 (1983).

\bibitem{metropolis}N.~Metropolis, A.W.~Rosenbluth, M.N.~Rosenbluth, A.H.~Teller, E.~Teller, J. Chem. Phys. {\bf 21}, 1087 (1953).

\bibitem{Will}W.~Lee, N.~Jean and S.~Sanvito, Phys. Rev. B  B {\bf 79}, 085120 (2009).

\bibitem{saxena}S.J.~Xie, K.H.~Ahn, D.L.~Smith, A.R.~Bishop and A.~Saxena, Phys. Rev. B {\bf 67}, 125202 (2003).

\end{thebibliography}
\end{document}